\documentclass[a4paper,11pt]{article}
\pdfoutput=1 
\usepackage{jheppub}

\usepackage{graphicx}
\usepackage{dcolumn}
\usepackage{bm}
\usepackage{epstopdf}
\usepackage{mathrsfs}
\usepackage{amssymb,amsmath,amsfonts,latexsym}
\allowdisplaybreaks
\usepackage{nicefrac}
\usepackage[colorlinks=true,linktocpage=true,linkcolor=blue,citecolor=blue]{hyperref}
\usepackage[utf8]{inputenc}
\usepackage{lipsum}
\usepackage{nicefrac}
\usepackage{slashed}
\usepackage{mathtools}
\def\bs{\boldsymbol} 
\def\del{\partial}
\def\bdel{\bs\partial}
\newcommand{\eqn}[1]{Eq.~\eqref{#1}}

\long\def\comment#1{ }

\def\and{\qquad\text{and}\qquad} 
\def\el{\text{el}} 
 
\def\eff{\text{eff}} 
\def\BH{\text{BH}}
\def\HO{\text{HO}} 
\def\fac{\text{sub}}

\def\mfp{\text{mfp}}

\def\mfp{\text{mfp}}

\def\glug{{\rm gg}}
\def\gq{{\rm gq}}
\def\qg{{\rm qg}}
\def\qq{{\rm qq}}

\def\q{{\rm q}}

\def\0{{\boldsymbol 0}}
\def\q{{\bm q}}

\def\x{{\boldsymbol x}}
\def\y{{\boldsymbol y}}

\def\z{{\boldsymbol u}}

\def\tform{t_\text{f}}

\def\qin{{\hat q_0}}

\def\HO{\text{HO}}

\def\Kc{{\cal K}}

\newcommand{\beq}{\begin{eqnarray}}
\newcommand{\eeq}{\end{eqnarray}}
\newcommand{\be}{\begin{eqnarray*}}
\newcommand{\ee}{\end{eqnarray*}}
\newcommand{\bal}{\begin{align}}
\newcommand{\eal}{\end{align}}
\newcommand{\rmd}{{\rm d}}
\newcommand{\dd}{{\rm d}}
\newcommand{\rme}{{\rm e}}

\def\rmR{{\rm Re}}

\newcommand{\nn}{\nonumber\\ }

\begin{document}

\title{Improved opacity expansion for medium-induced parton splitting
}

\author[a]{Yacine Mehtar-Tani} 
\author[b]{Konrad Tywoniuk}

\affiliation[a]{Physics Department, Brookhaven National Laboratory, Upton, NY 11973, USA}
\emailAdd{mehtartani@bnl.gov}

\affiliation[b]{Department of Physics and Technology, University of Bergen, 5007 Bergen, Norway}
\emailAdd{konrad.tywoniuk@uib.no}

\date{\today}
\abstract{

We present a new expansion scheme to compute the rate for parton splittings in dense and finite QCD media. In contrast to the standard opacity expansion, our expansion is performed around the harmonic oscillator whose characteristic frequency depends on the typical transverse momentum scale generated in the splitting. The first two orders account for the high frequency regime that is dominated by single hard scatterings together with the regime of multiple soft scatterings at low frequency.  This work generalizes the findings of Ref.~\cite{Mehtar-Tani:2019tvy} beyond the leading logarithmic approximation allowing to account also for the Bethe-Heitler regime and compare to the full numerical results from \cite{CaronHuot:2010bp}.  We investigate the sensitivity of our results to varying the separation scale that defines the leading order. Finally, the application to Monte Carlo event generators is discussed.

}

\date{\today}
\maketitle
\flushbottom

\section{Introduction}
\label{sec:intro}

Measurements of significant modifications of hard probes observables, in particular jets, in heavy ion collisions as compared to proton-proton collisions at RHIC and LHC have firmly established the prominent role of final-state interactions in the dense nuclear medium created in heavy-ion collisions.
For large systems, radiative processes are the main mechanism responsible for the observed quenching effects \cite{Peigne:2008wu} (for recent reviews see \cite{Mehtar-Tani:2013pia,Blaizot:2015lma}). Therefore a precise description of these processes is essential for a quantitative understanding of the mechanisms driving in-medium jet modification and probing non-equilibrium dynamics of the quark-gluon plasma (QGP).  

For energetic particles propagating close to the light-cone through a QCD medium, the problem reduces to describing the (non-relativistic) dynamics in the transverse plane perpendicular to the direction of propagation.
A formalism for dealing with multiple scattering in a QCD medium was developed by Baier-Dokshitzer-Mueller-Peign\'e-Schiff \cite{Baier:1996kr,Baier:1996sk} and Zakharov \cite{Zakharov:1996fv,Zakharov:1997uu}, usually referred to as the BDMPS-Z formalism.\footnote{See \cite{Arnold:2002ja} for an analogous approach within thermal field theory.} This resummation \cite{Wiedemann:2000za} can also be cast as an expansion in medium opacity \cite{Gyulassy:2000er,Ovanesyan:2011kn,Sievert:2018imd,Sievert:2019cwq}.\footnote{For a similar effort within the so-called higher-twist formalism, see \cite{Wang:2001ifa,Majumder:2009ge}.}

The regime of strong interactions can be approximated by diffusive broadening of the transverse momentum, governed by the diffusion coefficient $\hat q$. In this regime, characterized by the formation time of the radiation, $\tform$, being larger than the mean free path $\ell_\mfp$, i.e.  $\ell_\mfp \ll \tform \lesssim  L $, interference effects between subsequent scattering centers have to be taking into account leading to the Landau-Pomeranchuk-Migdal (LPM) suppression.
In this regime, the transport coefficient $\hat q$ is proportional to the Coulomb logarithm which must be regulated in the UV by the typical transverse momentum acquired by many soft scatterings, $q_{\rm med}^2$, see e.g. \cite{Arnold:2008zu}.  This approximation ceases to be valid for relatively hard emissions for which the dominant effect comes from single scattering with the medium  \cite{Zakharov:2000iz,Arnold:2009mr}.  Apart from a full numerical solution of the propagator \cite{CaronHuot:2010bp,Feal:2018sml,Ke:2018jem}, it is currently unclear how to account for both regimes in a semi-analytic fashion. 

The sensitivity of $\hat q$ to a high-energy cut-off scale comes from the underlying assumption that the medium in heavy-ion collisions consists of dressed quasi-particles whose interaction cross-section is given by a Coulomb-like power-law.
A first step towards the unification of the two limits described above was undertaken in \cite{Mehtar-Tani:2019tvy}. The main idea, inspired by the Moli\'ere theory of scattering \cite{Moliere:1948zz} (see also \cite{Iancu:2004bx} for a more recent application in the context of momentum broadening in high energy proton-nucleus collisions), is to treat the leading logarithm generated by the Coulomb tail in the medium potential to all orders in opacity since in this case the problem simplifies to solving for a harmonic oscillator potential and the remainder is treated a perturbation.  
The expansion parameter will thus be $\ln^{-1} (q_{\rm med}^2/\mu^2)$, where $q_{\rm med}$ the typical transverse momentum transfer with the medium and $\mu$ an IR cut-off such as the Debye mass. 

In this work, we generalize the approach in \cite{Mehtar-Tani:2019tvy} to go beyond the leading-logarithmic corrections to $\hat q$. This is achieved by expanding the scattering kernel around a harmonic oscillator approximation to incorporate the effects of hard scattering on top of multiple soft interactions. This is equivalent to a shift of the conventional opacity expansion around vacuum propagation to a solution that directly accounts for multiple soft scattering in the medium. Since the procedure in principle encodes the full information about the power-law behavior of Coulomb interactions, it also describes the Bethe-Heitler regime of very soft gluon emissions, i.e. $\tform \lesssim L$, which is especially relevant for dilute media, $\ell_{\rm mfp} \lesssim L$. Our result for the spectrum of medium-induced splittings is therefore valid for arbitrary energies and medium sizes, interpolating between three regimes: 1) Bethe-Heitler ($\tform < \ell_{\rm mfp}$), 2) LPM ($\ell_{\rm mfp} < \tform < L$) and 3) single, hard scattering ($L< \tform$). Although we implicitly assume that the jet is energetic enough such that $\sqrt{E/\hat q} > L$ (the ``thin'' medium limit, according to \cite{Arnold:2009mr}), our improved formula also accounts for ``thick'' media where $L> \sqrt{E/\hat q}$. 

The manuscript is organized as follows. In Sec.~\ref{sec:spectrum}, we recall the general expression for leading order splitting distribution in the presence of a dense QCD medium. In Sec.~\ref{sec:expanding}, we evaluate the spectrum by expanding close to the Harmonic Oscillator. We carry out the analytic calculations for the first two terms that are sufficient to interpolate between multiple soft and single hard approximations. In Sec.~\ref{sec:numerics}, our results for the spectrum and the rate are plotted. The latter is compared to the full numerical results from the McGill group which was first presented in \cite{CaronHuot:2010bp}.

\section{Spectrum of medium-induced gluons}\label{sec:spectrum}

The probability for a high energy parton $a$, of energy $E$, to split into a two partons $b$ and $c$ carrying a fraction $z$ and $1-z$ of its energy, respectively, due to multiple scatterings in a dense QCD medium is given by 
\begin{align}\label{eq:spectrum}
z\frac{\dd I_{ba}}{ \dd z} &= \frac{\alpha_s \, zP_{ba}(z)}{(z(1-z)E)^2} \, 2\rmR \int_0^\infty \dd t_2 \int_0^{t_2} \dd t_1 \,\nn
&\times  \bdel_\x \cdot \bdel_\y \Big[\Kc_{ba}(\x,t_2;\y,t_1) - \Kc_{0}(\x,t_2;\y,t_1) \Big]_{\x=\y=0}  \,.
\end{align}
where $P_{b(c) \leftarrow a}(z)$ stands for the Altarelli-Parisi splitting functions that read for the various relevant branching processes
\begin{eqnarray}\label{eq:split-def}
P_\glug(z)  &=& \frac{1}{2}~2C_{A} \frac{[1-z(1-z)]^2}{z(1-z)}\,\\
P_\qg(z) &=& \frac{1}{2}~2 N_f T_{F} \Big(z^2+(1-z)^2\Big)\;, \\
P_\gq(z) &=& \frac{1}{2}~C_{F} \frac{1+(1-z)^2}{z}\;, \\
P_\qq(z)  &=& \frac{1}{2}~C_{F} ~\frac{1+z^2}{(1-z)}\;. 
\end{eqnarray}
where $C_A=N_c$, $C_F=(N_c^2-1)/2N_c$ and $T_F=1/2$, with $N_c=3$ for SU(3). $N_f$ is the active number of quark flavors which we fix to 3 for our applications throughout.
%
Note that we have adopted the instant form notations for simplicity. The time variables stand in fact for light-cone time, i.e., $t =x^+$ and the the mass (energy) $E$ corresponds the the longitudinal light-cone momentum $p^+$. 

The Green's function $\Kc(\x,t_2;\y,t_1) $ accounts for the interactions with the medium taking place during the formation time of the splitting. It implicitly depends on the color representation of the in-coming and out-going partons and obeys the following Sch\"odinger equation,
\beq
\label{eq:eom-full-0}
\left[i \frac{\partial}{\partial t}  + \frac{\bdel^2}{2 z(1-z) E } + i v_{ba}(\x,t) \right] \Kc_{ba}(\x,t; \y,t_0) = i \delta(t-t_0) \delta(\x-\y) \,,
\eeq
where medium interactions are incorporated via the interaction Hamiltonian $H_\text{int} = iv(\x,t)$. The free part of the non-relativistic Hamiltonian in 2+1 dimensions is given by $H_0 = i\partial_t + \bdel^2/(2\omega)$ with a ``mass'' parameter given by $\omega\equiv z(1-z)E$ is solved by the following vacuum Green's function,
\beq
\Kc_0(\x,t; \y,t_0) = \frac{\omega}{2\pi i (t-t_0)} \exp \left[ i\frac{\omega (\x-\y)^2}{2(t- t_0)} \right] \,.
\eeq
Naturally, for $z<1$ we can identify $\omega$ with the energy of the emitted, soft daughter particle, $\omega \approx zE$. Thus, \eqn{eq:eom-full-0} describes the propagation through, and subsequent transverse broadening in a medium described by an imaginary three-body potential $i v_{ba}(\x,t)$, which is given by
\begin{align}
v^c_{ba}(\x,t) &= \frac{C_b+C_c-C_a}{2} \tilde v(\x,t) +\frac{C_c+C_a-C_b}{2}\tilde v(z \x,t) \nn
&+\frac{C_a+C_b-C_c}{2} \tilde v\big((1-z)\x,t \big) \,,
\end{align}
where $C_{a}$,  $C_{a}$ and $C_{c}$ are the color factors associated with partons in representations $a$, $b$ and $c$, respectively. Explicitly, we have
\begin{align}
\label{eq:gluon-gluon-potential}
v_{\text{gg}}(\x,t) &= \frac{N_c}{2} \tilde v(\x,t) +\frac{N_c}{2} \tilde v(z \x,t) + \frac{N_c}{2} \tilde v\big((1-z)\x,t \big)\,,\\
v_{\text{gq}}(\x,t) &= \frac{N_c}{2} \tilde v(\x,t) +\left(C_F - \frac{N_c}{2} \right) \tilde v(z \x,t) + \frac{N_c}{2} \tilde v\big((1-z)\x,t \big)\,,\\
v_{\text{qq}}(\x,t) &= \frac{N_c}{2} \tilde v(\x,t) + \frac{N_c}{2} \tilde v(z \x,t) + \left(C_F - \frac{N_c}{2} \right) \tilde v\big((1-z)\x,t \big) \,,\\
v_{\text{qg}}(\x,t) &= \left(C_F - \frac{N_c}{2} \right) \tilde v(\x,t) + \frac{N_c}{2} \tilde v(z \x,t) + \frac{N_c}{2} \tilde v\big((1-z)\x,t \big) \,.
\end{align}
where\footnote{Throughout, we adopt the shorthand $\int_{\q}\equiv \int \dd^2 \q/(2\pi)^2$ and $\int_{\x} \equiv \int \dd^2 \x$ for the transverse integrals.} the potential is given by
\beq\label{eq:x-GW-potential}
\tilde v (\x,t) = \int_\q\, \, \frac{\rmd\sigma_\text{el}}{\rmd^2 \q} \left(1-\rme^{i\q\cdot\x}\right) \,.
\eeq
The elastic scattering potential with the medium can be extracted from thermal field theory in a weakly-coupled medium \cite{Aurenche:2002pd,Arnold:2003zc,CaronHuot:2008ni}, but is  often modeled as
\beq \label{eq:el-GW}
 \frac{\rmd^2 \sigma_\text{el}}{\rmd^2 \q} \equiv \frac{g^4 n}{(\q^2 + \mu^2)^2},
\eeq
also referred to the Gyulassy-Wang model \cite{Wang:1991xy} (here, $n= n(t) \sim T^{-3}$ corresponds to the density of scattering centers in the medium). The potential is screened at the scale $\mu$ that is related to the Debye mass in a thermal medium, i.e., $\mu^2 \sim m_D^2 = (1+N_f/6) g^2T^2$.

In the case of interest many scattering centers contribute during the branching process and as a result the typical transverse momentum acquired by the three body system $(a,b,c)$ is much larger than the Debye mass $k_\perp\sim x^{-1}_\perp \gg \mu $. As a result, $\tilde v (\x,t)$ is dominated by a large Coulomb  logarithm $\ln (x_\perp \mu)^{-1}$. Using the GW model for the potential given by \eqn{eq:el-GW}, we find
\begin{align}\label{eq:potential-log}
\tilde v (\x,t) &= \int_\q\, \, \frac{\rmd\sigma_\text{el}}{\rmd^2 \q} \left(1-\rme^{i\q\cdot\x}\right)= \frac{\hat q_0(t)}{\mu^2N_c} \big[1 - \mu |\x| K_1( \mu |\x|) \big]  \,,\\
&\approx \frac{1}{4N_c} \x^2 \hat q_0(t) \left(\ln \frac{4}{\mu^2 \x^2} +1 - 2 \gamma_E \right) +\ldots
 \end{align}
up to sub-leading corrections of order $\sim \x^4$. Here, $K_1( x)$ is the modified Bessel function of the second kind and $\gamma_E \approx 0.577\ldots$ the Euler-Mascheroni constant. 
For later convenience we have introduced the  ``bare'' quenching parameter, stripped of the Coulomb logarithm 
\beq\label{eq:qhat-stripped}
\qin(t)\equiv  4\pi \alpha_s^2 N_c \, n(t). 
\eeq 
For a thermal medium we simply have $n=(3/2) T^3$. For $T=0.4$ GeV and $g=1.94$ (so that $\alpha_s \approx 0.3$) we find $\qin(t)=1.83$ GeV$^2$/fm and $m_D = 0.9$ GeV.

\section{Expanding around the harmonic oscillator}
\label{sec:expanding}

It is in general difficult to solve \eqn{eq:eom-full-0} exactly besides using numerical methods \cite{CaronHuot:2008ni,Feal:2018sml,Ke:2018jem}. A first strategy consists in a plain expansion in $v(\x,t)$, which stands for the standard opacity expansion, where opacity is defined as a ratio between the medium length and the mean free path $\sim \ell_{\rm mfp}/L$. The first order in the latter approach is typically referred to as the Gyulassy-Levai-Vitev (GLV) approximation \cite{Gyulassy:2000er}, see also \cite{Wiedemann:2000za}. However, this approximation breaks down in a dense medium at frequencies $\omega <\omega_c \sim \hat q L^2$, which for realistic values such as  $\hat q=2$ GeV$^2$/m and $L=5$ fm for instance yields a large value $\omega_c=250$ GeV. In this case, the transverse momentum accumulated during the branching is determined by multiple soft scattering, i.e. $k^2_\perp\sim \sqrt{\hat q \omega} $. In this non-perturbative regime the potential can be approximated by $v(\x,t) \sim \x^2$ by neglecting the variation of the Coulomb logarithm (cf.~  \eqn{eq:potential-log}).  In this case, the equation of motion is identical to that of a harmonic oscillator  with complex frequency, hence this scheme is often referred to as the ``harmonic oscillator'' (HO) approximation. Of course, in order to obtain a quantitively sound result one needs to estimate the argument of the logarithm which introduces an uncertainty which is of order the inverse of the Coulomb logarithm.  

Our strategy in what follows is to shift the expansion point to the ``harmonic oscillator'' solution as follows
\beq
v(\x,t) = v_\HO(\x,t) + \delta v(\x,t) \,,
\eeq
where $\delta v (\x,t) = v(\x,t) - v_\HO(\x,t) \ll v_\HO(\x,t)$,
 will be treated as a perturbation. The HO potential is given by
\beq
\label{eq:vho-1}
v_\HO(\x,t) \equiv \frac{1}{4} \x^2 \, \hat q_\eff(t)\,,
\eeq
with the effective $\hat q$ parameter chosen to be
\beq\label{eq:qeff}
\hat q_\eff(Q^2_\fac,t) = \frac{1}{2}  \left[1 + \left(\frac{2 C_R}{N_c} -1 \right) z^2 + (1-z)^2\right]\, \hat q_0(t)\ln \frac{a Q^2_\fac}{\mu^2} \,.
\eeq
where $\hat q_0(t)$ is given in \eqn{eq:qhat-stripped} and $a=4 \rme^{-2\gamma_E+1}$. Here, we have limited our discussion to a parton in representation $R=q,g$ that radiates a gluon of energy fraction $z$. The constant $a$ accounts for the constant terms in \eqn{eq:potential-log}.

 The jet quenching coefficient $\hat q_\eff$ is logarithmically dependent on a subtraction scale which is only a function of $\omega$, i.e. $Q_\fac = Q_\fac(\omega)$. This scale must be chosen to be the typical transverse momentum generated during the splitting, that is, 
 \beq\label{eq:sub-scale}
 Q^2_\fac\simeq \sqrt{z(1-z) E \hat q_\eff (Q^2_\fac)}\, 
 \eeq
 in the HO approximation. 
 
 The advantage of choosing this expansion point, is that in the HO approximation the Sch\"odinger equation,
 \beq
\label{eq:eom-HO}
\left[i \frac{\partial}{\partial t}  + \frac{\bdel^2}{2 \omega } + i v_\HO(\x,t) \right] \Kc_\HO(\x,t; \y,t_0) = i \delta(t-t_0) \delta(\x-\y) \,,
\eeq
admits the known analytic solution \cite{Abramowitz},
\beq\label{eq:green-fct-sol}
\Kc_\HO(\x,t ; \y ,t_0) = \frac{i \omega}{2 \pi S(t,t_0)} \exp \left\{ i \frac{\omega}{2 S(t,t_0)} \left[C(t_0,t)\, \x^2 + C(t,t_0) \,\y^2 - 2 \x\cdot \y \right] \right\} \,.
\eeq
Here, the functions $S(t,t_0)$ and $C(t,t_0)$ represent the two independent solution of the equation
\beq
\frac{\dd^2 f(t)}{\dd t^2} = - \Omega^2(t) f(t) \,,
\eeq
where the frequency is given by $\Omega(t) = \sqrt{\hat q(t)/(2i\omega)}$, with boundary conditions $S(t_0,t_0) = 0$ and $\partial_t S(t,t_0) \vert_{t=t_0} = 1$, and $C(t_0,t_0) = 1$ and $\partial_t C(t,t_0) \vert_{t=t_0}=0$, respectively. They are related through a constant Wronskian,
\beq
W = C(t,t_0) \partial_t S(t,t_0) - S(t,t_0) \partial_t C(t,t_0) \,.
\eeq
Given the boundary conditions on the functions $S$ and $C$, it turns out that $W=1$.
For a medium of constant density and length $L$, where $\Omega(t) = \Omega$, we find that $S(t,t_0) = \sin[\Omega(t-t_0)]/\Omega$ and $C(t,t_0) = \cos[\Omega(t-t_0)]$.

The full solution can then be recast as an implicit equation, and reads
\begin{align}
\Kc(\x,t_1; \y,t_0) &= \Kc_\HO (\x,t_1; \y,t_0) \nn
&- \int \dd^2 \z \int_{t_0}^{t_1} \dd t \, \Kc_\HO (\x,t_1; \z,t) \delta v(\z,t) \Kc(\z,t ; \y,t_0 ) \,.
\end{align}
In what follows we shall solve the latter iteratively for the first two orders: the leading-order (LO) term reads $\Kc_{\rm LO} = \Kc^{(0)}$, where
\beq
\label{eq:nlo-general-0}
\Kc^{(0)}(\x,t_1; \y,t_0)  = \Kc_\HO (\x,t_1; \y,t_0) \,.
\eeq
The next-to-leading (NLO) correction is given by
\beq
\label{eq:nlo-general-1}
\Kc^{(1)}(\x,t_1; \y,t_0) = - \int \dd^2 \z \int_{t_0}^{t_1} \dd t \, \Kc_\HO (\x,t_1; \z,t) \delta v(\z,t) \Kc_\HO (\z,t ; \y,t_0 ) \,,
\eeq
so that the full NLO term is simply $\Kc_{\rm NLO} = \Kc^{(0)} + \Kc^{(1)}$. Higher orders are found in a analogous manner. Our results below show that the two first terms already provide a reasonable description of the spectrum and the related splitting rate.

\subsection{Leading order: the harmonic oscillator approximation} 

Let us first consider the leading order that corresponds the BDMPS approximation. Inserting \eqn{eq:green-fct-sol} in \eqn{eq:spectrum} yields
\beq\label{eq:spectrum-HO}
z\frac{\rmd I^{(0)}}{\rmd z}  \equiv z\frac{\rmd I_\HO}{\rmd z} = \frac{\alpha_s zP(z)}{\pi} 2\rmR  \int_0^\infty \rmd t_2 \int_0^{t_2}\rmd t_1 \,  \, \left[\frac{1}{S^2(t_2,t_1)} - \frac{1}{(t_2-t_1)^2}\right] \,,
\eeq
where we have left the indices indicating the parton splitting to be implicit.
Using the following property \cite{Arnold:2008iy}
\beq\label{eq:dcot}
\del_t \left(\frac{C(t,s)}{S(t,s)}\right) =-\frac{1}{S^2(t,s)}\,,
\eeq
the $t_2$ integration can be carried out and reads
\beq\label{eq:HO-t2-integral}
 \int_{t_1}^\infty \rmd t_2 \, \frac{1}{S^2(t_2,t_1)} = \frac{C(t_1,t_1)}{S(t_1,t_1)}-\frac{C(\infty,t_1)}{S(\infty,t_1)}\,.
\eeq
The first term in \eqn{eq:HO-t2-integral} cancels against the vacuum piece, i.e. the second term in \eqn{eq:spectrum-HO},  while the second one can be integrated further over $t_1$, 
\beq\label{eq:HO-t2-integral-2}
 \int_{0}^\infty \rmd t_1 \frac{C(\infty,t_1)}{S(\infty,t_1)} =-  \int_{0}^\infty \rmd t_1 \frac{\del_{t_1}C(t_1,L)}{C(t_1,L)}  = \ln C(0,L) = \ln \cos(\Omega L)\,,
\eeq
where we have used the decomposition of $C(\infty,s)$ and $S(\infty,s)$ as a superposition of other solution to the wave equation \cite{Arnold:2008iy},
\begin{align}
S(t,t_1) &=C(t_1,t_0) S(t,t_0)-S(t_1,t_0) C(t,t_0)\,,\nn
C(t,t_1) &=-\del_{t_1}C(t_1,t_0) S(t,t_0)-\del_{t_1}S(t_1,t_0) C(t,t_0)\,.\nonumber
\end{align}
Hence, letting $t=\infty$, $t_1=s$ and $t_0=L$ yields
\begin{align}
\label{eq:S-C-composition}
S(\infty,s) &=C(s,L) S(\infty,L)-S(s,L) C(\infty,L) \,,\\
C(\infty,s) &=-\del_{s}C(s,L) S(\infty,L)+\del_{s}S(s,L) C(\infty,L) \,.
\end{align}
Finally, inserting \eqn{eq:HO-t2-integral-2} into \eqn{eq:spectrum-HO}  yields the BDMPS-Z result 
\beq\label{eq:bdmps-1}
  z\frac{\rmd I^{(0)}}{\rmd z} = \frac{2\alpha_s}{\pi} \,z P(z)\,  \ln |\cos(\Omega L)| \,.
\eeq
\eqn{eq:bdmps-1} encompasses two regimes
\beq
z\frac{\rmd I^{(0)}}{\rmd z} \simeq   \frac{\alpha_s}{\pi} \,z P(z)
 \begin{dcases}
\,\,\sqrt{\frac{\omega_c}{2\omega}}   \quad\qquad\text{for}\quad \omega \ll \omega_c\\
\,\,\frac{1}{12} \left(\frac{\omega_c}{\omega} \right)^2\quad\text{for} \quad \omega \gg \omega_c\\
\end{dcases}
\eeq
expressed in terms of the characteristic frequency 
\beq
\omega_c =\frac{1}{2}\hat q_\eff L^2\,.
\eeq
Recall that $q_\eff $ is a function of $z$ and depends on the flavor of the partonic configuration (cf. \eqn{eq:qeff}).
\subsection{Next-to-leading order correction to the harmonic oscillator} 

Let us turn now to evaluating  the next-to-leading term, given as a sum of Eqs.~(\ref{eq:nlo-general-0}) and (\ref{eq:nlo-general-1}). The physical meaning of the NLO correction is that one ``soft'' scattering, described purely via diffusive transverse momentum broadening, is replaced by a ``hard'' scattering, i.e. an interaction with the medium described by the Coulomb potential. This opens for the possibility that a single, hard kick from the medium can dominate the total transverse momentum transversed from the medium during the formation time. The correction to the splitting distribution reads, see \cite{Mehtar-Tani:2019tvy} for more details,
\beq\label{eq:spectrum-general-1}
z \frac{\dd I^{(1)}}{\dd z} = \frac{\alpha_s zP(z)}{\pi^2} 2 \rmR \int_0^L \dd s \int \frac{\dd^2 \z}{\z^2} \, \delta v(\z,s) \rme^{- k^2(s) \z^2}\,,
\eeq
where the indices are suppressed and
\beq
k^2(s) = i\frac{\omega}{2} \left[\frac{C(0,\infty)}{S(0,\infty)} - \frac{C(\infty,s)}{S(\infty,s)} \right] \,.
\eeq
In particular, for a medium with constant density $n(s) = n \Theta(L-s)$, we find
\beq\label{eq:ks}
k^2(s)= i  \frac{\omega \Omega}{2} \big[ \cot\big(\Omega s\big)-  \tan\big(\Omega(L-s) \big) \big]\,.
\eeq
For simplicity and without loss of generality, let us focus on the case where a gluon of energy fraction $z$ is emmitted of a parton $R$. Then, 
\beq \label{eq:delta-v}
\delta v_{\text{gR}}(\x,t) &=& \frac{N_c}{2} \delta \tilde v(\x,t)+\left(C_R - \frac{N_c}{2} \right) \delta\tilde v(z \x,t) + \frac{N_c}{2} \delta\tilde v\big((1-z)\x,t \big)\,,
\eeq
where
\beq
\delta \tilde v(\x,t) =  \int_{\q} \, \frac{\dd^2 \sigma_\el}{{\dd  \q}^2} \left(1-\rme^{i \q \cdot \z}\right) -  \frac{1}{4} \z^2  \hat q_0 \ln\frac{Q_\fac^2}{\mu^2} 
\eeq
In order to integrate over $\z$ it is convenient to use the Fourier representation of the dipole cross-section. Consider for instance the contribution from the second term in \eqn{eq:delta-v},
\begin{align}
\label{eq:spec-corr-a}
z \frac{\dd I^{(1)}}{\dd z}\bigg|_a &=  \frac{\alpha_s zP(z)}{\pi^2}\,\left(C_R - \frac{N_c}{2} \right) \, 2 \rmR \int_0^L \dd s  \int \frac{\dd^2 \z}{\z^2}  \int_{\q} \, \frac{\dd^2 \sigma_\el}{{\dd  \q}^2} \left(1-\rme^{i z \q \cdot \z}\right)\,\rme^{- k^2(s) \z^2} \,,\\
\label{eq:spec-corr-b}
z\frac{\dd I^{(1)}}{\dd z}\bigg|_b &=  - \frac{\alpha_s zP(z)}{\pi^2}\,\,\left(C_R - \frac{N_c}{2} \right)  2 \rmR \int_0^L \dd s  \int \dd^2 \z \, \frac{1}{4}\, z^2 \, \hat q_0 \, \rme^{- k^2(s) \z^2} \,.
\end{align}
The contribution $a$ relates to the compete dipole cross-section from which the HO part, i.e. contribution $b$, must be subtracted.
Strikingly, the integrations can be performed analytically by noticing that the $u$ integral yields
\beq
\int_0^\infty \frac{\rmd  u}{u}\, \big[ 1-J_0(z q u ) \big] \, \rme^{-u^2 k^2(s)} = \frac{1}{2}\left[ \gamma_E+\Gamma\left(0,\frac{z^2q^2}{4k^2(s)}\right) +\ln\left(\frac{z^2q^2}{4k^2(s)}\right)\right] \,.
\eeq
Furthermore, by changing variables to $y =z^2 q^2/(4 k^2(s))$ we find
\beq
\int \frac{\rmd y}{(y+x)^2} \,\left( \gamma_E+\Gamma (0,y) -\ln y \right) = \frac{1}{x} \left[\gamma_E+\rme^{x}\Gamma\left(0, x\right)+\ln x\right],
\eeq
where $x = z^2\mu^2/[4 k^2(s)]$.
Finally, the second term of \eqn{eq:delta-v} becomes
\begin{align}
z\frac{\dd I^{(1)}}{\dd z} &=  \frac{\alpha_s zP(z)}{\pi} \left(\frac{2C_R}{N_c} -1 \right)  \, 2\rmR \int_0^L \dd s \nn
& \times \left\{\,\frac{ \hat q_0}{2\mu^2} \, \left[\gamma_E + \rme^{z^2\mu^2/[4k^2(s)]} \Gamma \left(0,\frac{z^2\mu^2}{4 k^2(s)}\right) + \ln \frac{z^2\mu^2}{4 k^2(s)} \right] 
- \frac{z^2  \hat q_0}{8 k^2(s)} \ln \frac{Q^2_\fac}{\mu^2} \right\}
\end{align}
where the first term appears due to the full elastic cross section, cf. \eqn{eq:spec-corr-a}, while the second term is the subtraction of the harmonic oscillator term, cf. \eqn{eq:spec-corr-b}. The first and third remaining terms can be found by simply substituting $z\to 1$ and $z \to 1-z$, respectively. Explicitly, the full NLO correction then takes the form
\begin{align}
\label{eq:bdmps-2}
z \frac{\dd I^{(1)}}{\dd z} &= \frac{\alpha_s \, zP(z)}{\pi} 2 \rmR \int_0^L \dd s\, \frac{\hat q_0}{2 \mu^2} \nn
&\times\left\{F\left(\frac{\mu^2}{4 k^2(s)} \right) +\left(\frac{2C_R}{N_c} - 1 \right) F\left(\frac{z^2\mu^2}{4 k^2(s)} \right) + F\left(\frac{(1-z)^2\mu^2}{4 k^2(s)} \right) \right\} \nn
& - \frac{\alpha_s \, z P(z)}{\pi} 2\rmR \int_0^L \dd s \,\frac{\hat q_\eff}{4 k^2(s)} \ln \frac{Q_\fac^2}{\mu^2} \,,
\end{align}
where we introduced the shorthand
\beq
F(x) \equiv \gamma_E+ \rme^x \Gamma(0,x) + \ln x \,.
\eeq
The full spectrum at NLO therefore becomes the sum of Eqs.~(\ref{eq:bdmps-1}) and (\ref{eq:bdmps-2}).
 
 Let us investigate the two limits of the complex function $k^2(s)$, given in \eqn{eq:ks}. In the limit $\omega \gg \hat q L^2$, it follows that $\Omega \ll 1 $, leading to $\tan(\Omega(L-s))\approx 0$ and $\cot\Omega s \approx (\Omega s)^{-1}$. It follows that
\beq
 k^2(s) \simeq  i \frac{\omega}{2s} \,,
\eeq
as in the vacuum. 
Then, expanding for small $x = \mu^2/[4 k^2(s)]$, we find
\beq
\frac{1}{x}\left[ \gamma_E + \rme^{x}\Gamma\left(0, x \right) + \ln x\right] \approx 1-\gamma_E + \ln\frac{1}{x} \,.
\eeq
Hence, the spectrum in the high frequency regime $\omega \gg \hat q L^2$ becomes
\beq
\omega \frac{\dd I}{\dd \omega} \approx \bar \alpha \, \frac{\pi}{4} \frac{\hat q L^2}{\omega} \,,
\eeq
which is a well-known limit of the GLV spectrum.

Turning to the small frequency regime, note that $k^2(s)$ also becomes vacuum-like since $Q_\fac^2 \approx \mu^2$ for $\omega < \omega_\BH$ and therefore $\hat q \to 0$. We can now expand for {\it large} $x$, to find
\beq
\omega \frac{\dd I}{\dd \omega} \approx 2 \bar \alpha \frac{\hat q_0 L}{\mu^2} \left(\ln \frac{\mu^2 L}{2 \omega} -1 + \gamma_E \right) \,.
\eeq
This is the characteristic behavior of the Bethe-Heitler regime, that is also contained in the GLV spectrum.

It is worth pointing out that these limits are universal and do not depend on the choice of matching scale $Q_\fac$ in $\hat q$. The approach to these values is however affected by the exact value of the matching scale, which we shall explore in the next section. 
%
%
%

\section{Numerics}
\label{sec:numerics}

Let us first return to the subtraction scale $Q_\fac$ introduced earlier in the context the transport coefficient $\hat q = \hat q_0 \ln a Q_\fac^2/\mu^2$. 
As discussed in \cite{Mehtar-Tani:2019tvy,CaronHuot:2008ni}, it is natural to define it in relation to the  characteristic transverse momentum of the medium-induced emission process, i.e. $Q_\fac^2 \sim k_\perp^2$. For radiative processes, we have that $k_\perp^2 \sim (\omega \hat q )^{1/2} \sim (\omega \hat q_0 \ln\left(a \sqrt{\omega \hat q_0}/\mu^2 \right))^{1/2}$, since $\hat q$ itself is running with the subtraction scale.\footnote{Other choices are indeed possible, e.g. fixing $Q_\fac^2 \sim \hat q L \sim \hat q_0 L \ln \left(\hat q_0 L/\tilde \mu^2 \right)$, which is the expected behavior at high energies $\omega > \hat q L^2$. Alternatively, we could also demand that the scale saturates, i.e. $Q_\fac^2 \leq \hat q L$.} Crucially, to ensure the matching with the Bethe-Heitler spectrum at small gluon frequencies the latter logarithm should vanish for $Q^2 < \mu^2$. This implies that $\hat q \to 0$ when $\sqrt{\omega \hat q} <\mu^2$ or $\omega < \omega_{BH} \equiv \mu^2 \ell_{\rm mfp}$ (where we used that $\hat q \sim  \mu^2/\ell_{\rm mfp}$ and we are left with a vacuum spectrum in this regime.
To do so we use the following interpolating form 
\beq
Q_\fac^2(\omega,\mu) = \sqrt{\omega \hat q_0 \ln\left(\frac{a \sqrt{\omega \hat q_0}}{\mu^2} \right)}\,  \rme^{-\mu^2/\sqrt{\hat q_0 \omega}}+\mu^2 \,.
\eeq
This choice guarantees that $Q_\fac$ smoothly goes to $ \mu$. Finally, in order to test the sensitivity to the chosen matching, we will multiply the numerical factor $a$ by $1/2$ and $2$.

\begin{figure}
\centering
\includegraphics[width=0.75\textwidth]{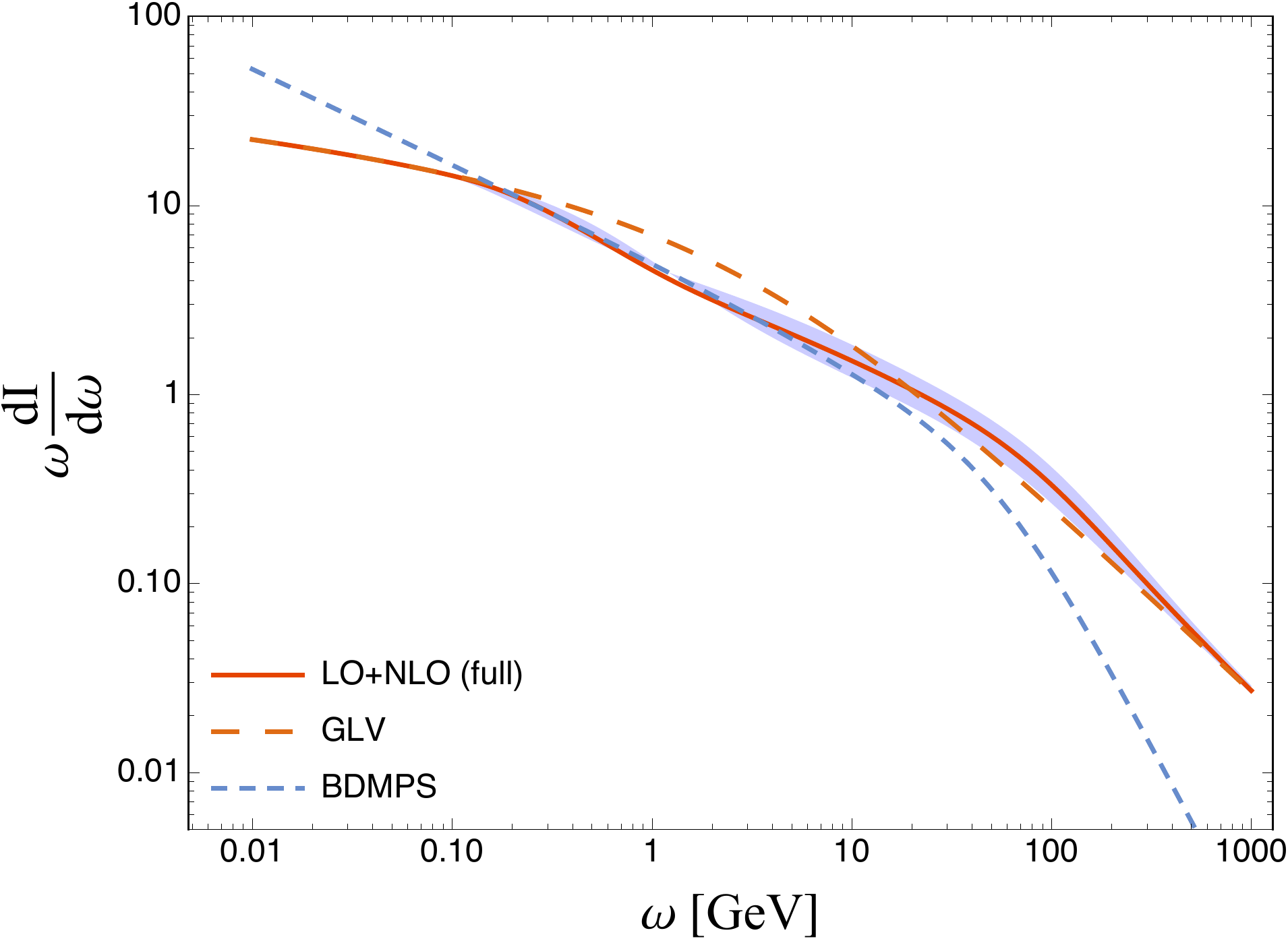}
\caption{The eikonal spectrum (assuming $x \ll 1$) up to next-to-leading order, compared to GLV and BDMPS for a jet with energy $E=1000$ GeV. Medium parameters are $\hat q=1.5$ GeV$^2$/fm, $\mu = 1$ GeV and $L = 4$ fm.}
\label{fig:spectrum-comparison}
\end{figure}

Let us first turn to the {\it eikonal} limit, where assume $x \ll 1$.
We compare the results from our formula in the eikonal limit with the expectation from first order in opacity (for short, labelled ``GLV'') and the multiple soft scattering approximation (labelled ``BDMPS'') in \autoref{fig:spectrum-comparison}.
The spectrum interpolates well between the three physical regimes for in-medium QCD bremsstrahlung. For the choice of parameters here, corresponding to $\ell_{\rm mfp} < L < \sqrt{E/\hat q}$, it clearly demonstrates that LPM interference effects are suppressing the spectrum over a large range of gluon energies. It is worth keeping in mind that we have only included the first order correction to the standard HO baseline, which leaves further room to improve on the matching by adding higher orders. The band around the ``LO+NLO'' curve corresponds to varying the matching parameter $a$ by a factor 2 up and down, which describes the inherent ambiguity in defining the LPM suppressed regime. Most importantly, our results reproduce the universal expectations at low- (corresponding to the Bethe-Heitler limit) and the high-frequency (corresponding to the ``GLV'' limit) and, all in all, the result over a wide frequency range is quite good. 

\begin{figure}
\centering
\includegraphics[width=0.75\textwidth]{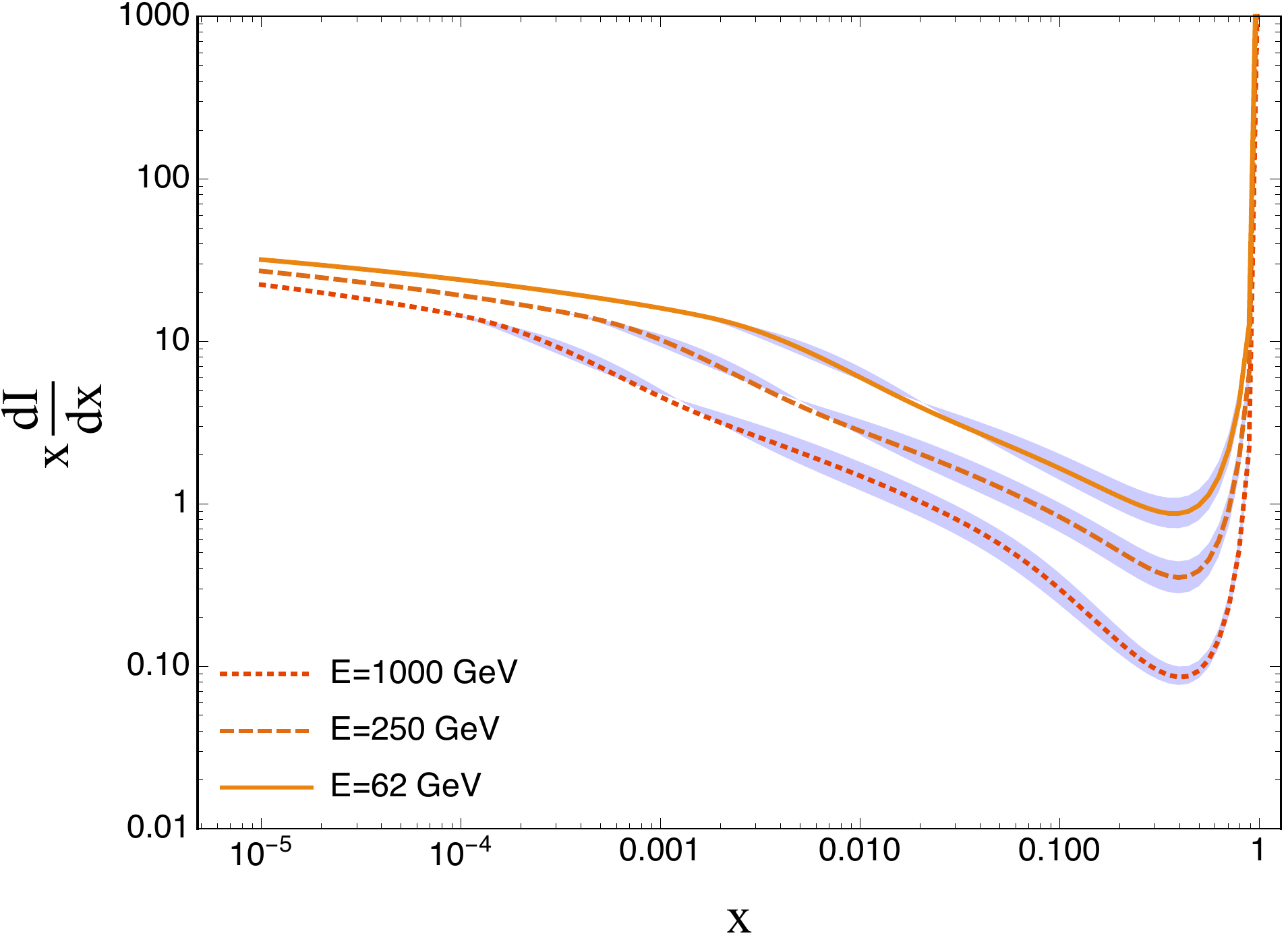}
\caption{The matched spectrum up to next-to-leading order, compared to GLV and BDMPS for a jet with energy $E = 250$ GeV (top) and $E=1000$ GeV (bottom). Medium parameters are $\hat q=1.5$ GeV$^2$/fm, $\mu = 1$ GeV and $L = 4$ fm.}
\label{fig:spectrum-1}
\end{figure}

The results for the improved spectrum with full $x$ dependence, calculated for a gluon jet with for three different energies $E=\{1000,\, 250, \,62.5\}$ GeV, are given in \autoref{fig:spectrum-1}. 
For the upper energy, the jet traverses a ``thin'' medium where $\sqrt{E/\hat q}> L$ while for the lower energy, the medium is ``thick'', i.e. $\sqrt{E/\hat q} < L$. This affects the range where the LPM is at play.
The turning over of the curves at small-$x$ correspond to the BH regime which takes place at a fixed gluon energy $\omega_{\rm BH}$, not a fixed $x$ for different jet energies.

\begin{figure}
\centering
\includegraphics[width=0.8\textwidth]{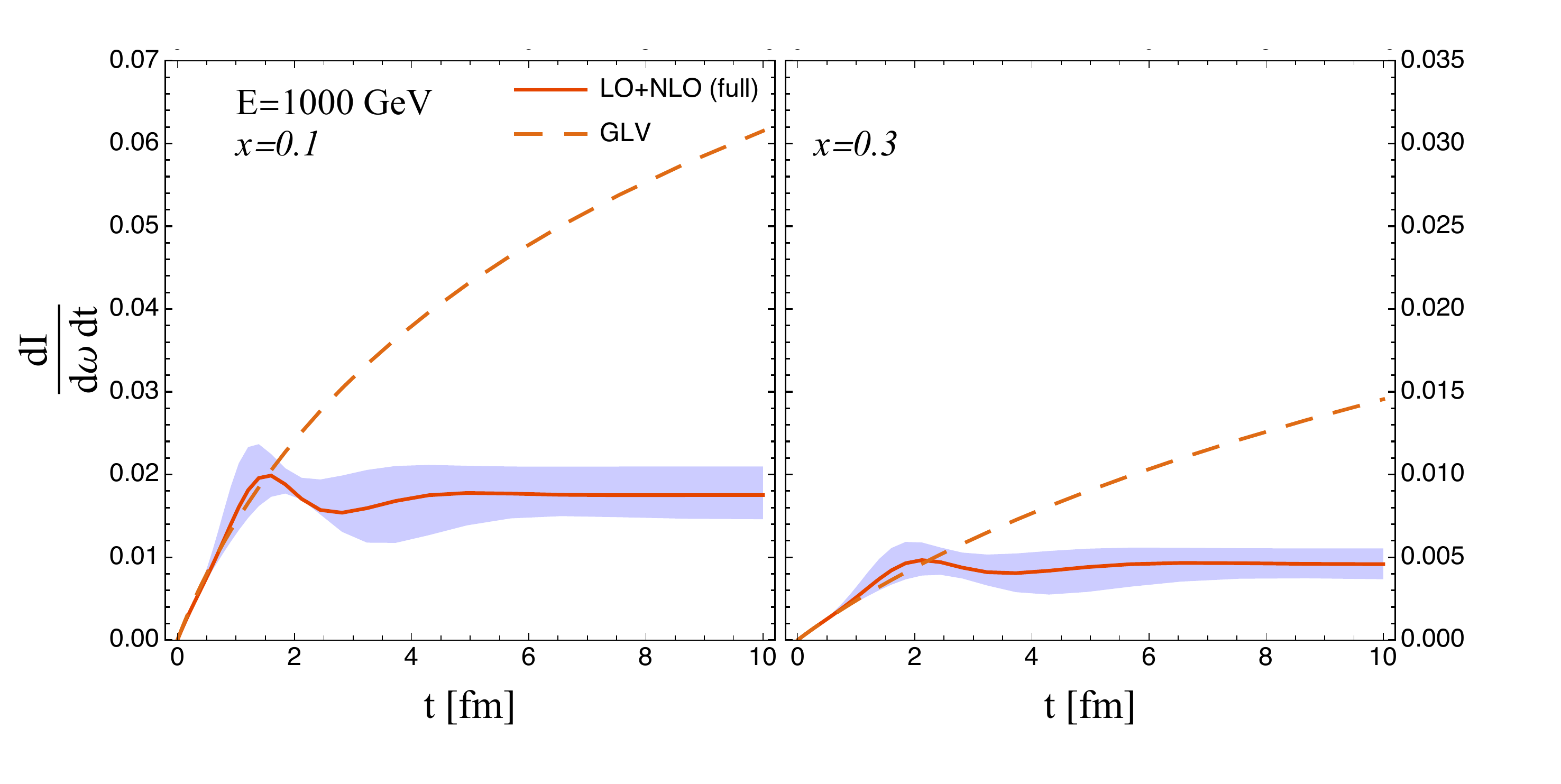}
\caption{The rate of medium-induced emissions for a gluon jet with energy $E = 1$ TeV. Medium parameters are $\hat q=1.5$ GeV$^2$/fm and $\mu = 1$ GeV.}
\label{fig:rate-1}
\end{figure}

We have also computed numerically the rate of emissions, defined as $\dd I/(\dd \omega \, \dd t)$, in \autoref{fig:rate-1}.
As expected, at small times the rate grows linearly with time like the GLV spectrum, i.e. $\propto t$. At later times the rate satures in the LPM regime. We observe the uncertainty related to the implementation of the LPM regime as a band that predominantly appears whenever the rate saturates. For the same reason, we have therefore chosen not to plot the expectation from the ``bare'' BDMPS spectrum since it would differ from our curve by a constant offset related to choice of scale in $\hat q$.

\begin{figure}
\centering
\includegraphics[width=0.8\textwidth]{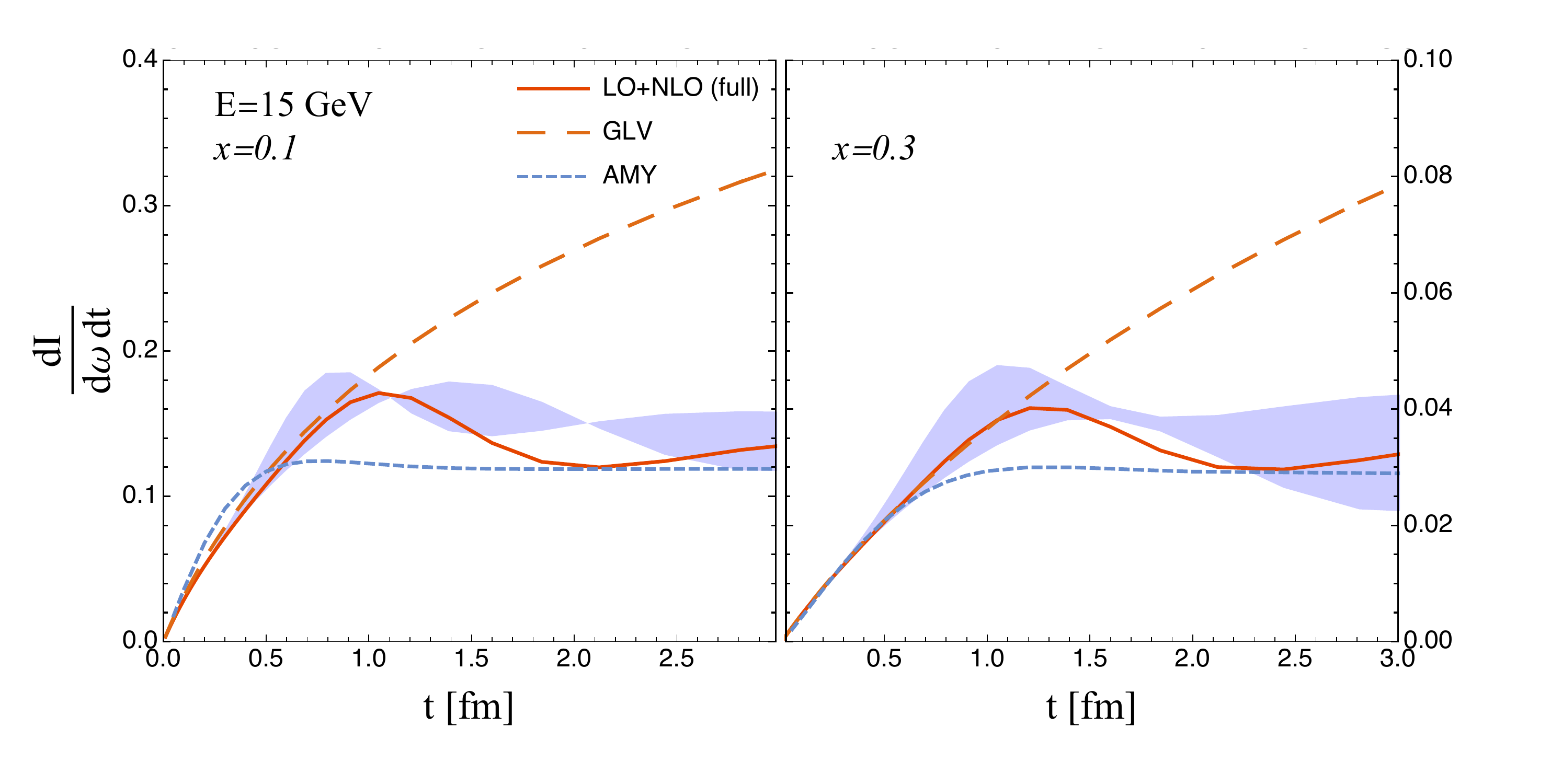}\\
\includegraphics[width=0.8\textwidth]{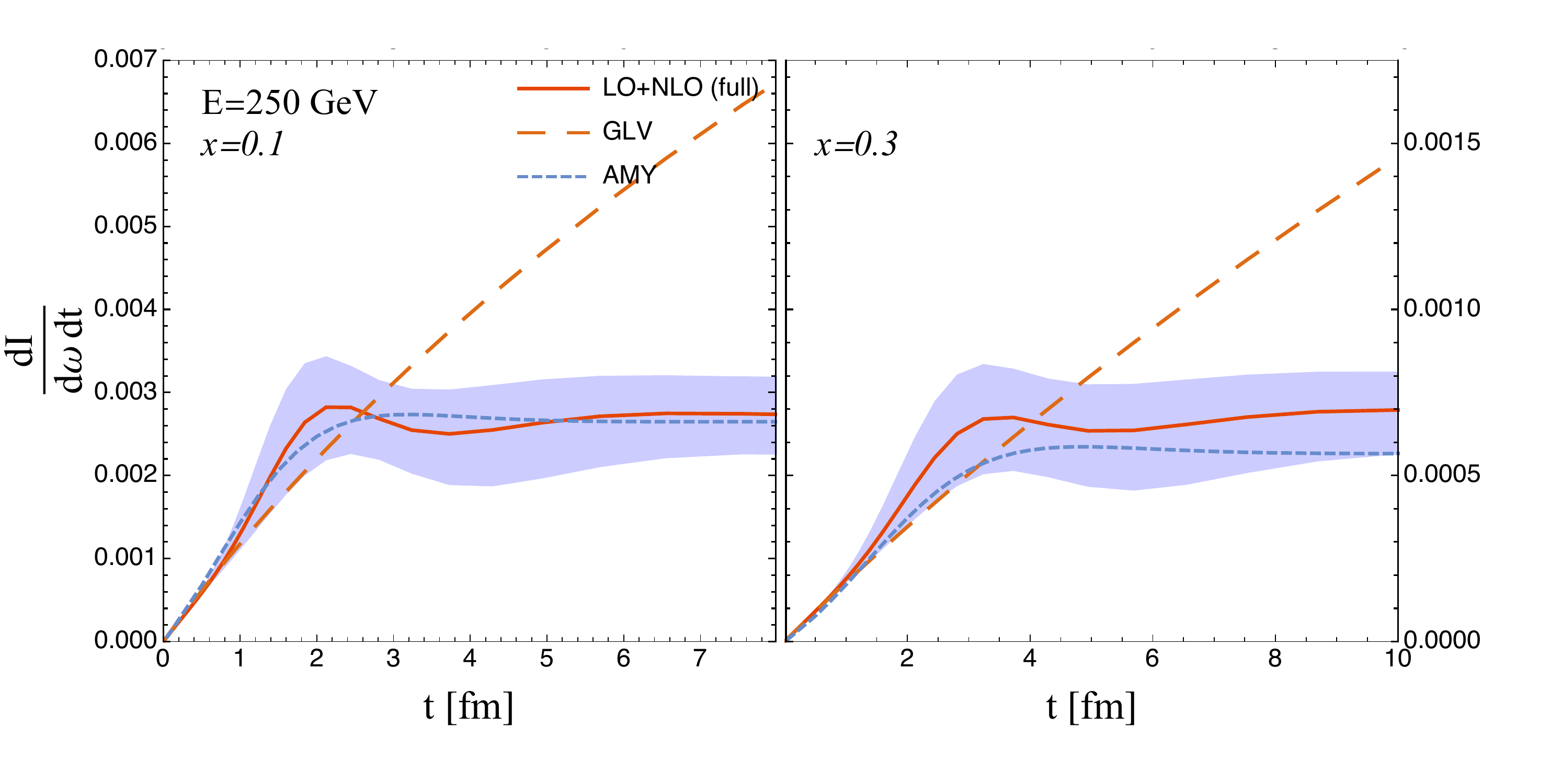}
\caption{The rate of medium-induced emissions for a jet with energy $E = 15$ GeV (upper panel) and $E = 250$ GeV (lower panel). Medium parameters are $\hat q=1.63$ GeV$^2$/fm and $\mu = 0.9$ GeV, corresponding to $g = 1.94$ and $T=0.4$ GeV. The dashed (blue) lines labelled ``AMY'' correspond to the numerical evaluation presented in \cite{CaronHuot:2010bp}.}
\label{fig:rate-2}
\end{figure}

Finally, in order to compare our compact, analytical formula with the full, all-order in opacity solution of the spectrum, solved numerical in \cite{CaronHuot:2010bp}, we plot in \autoref{fig:rate-2} the rate for two choices of medium parameters and jet kinematics.
Since our result only includes the first corrections from hard scattering, the agreement is reasonable ($\lesssim 30 \% $). Note that some of the discrepancy may be attributed to the different choices of the potential and the lack of thermal masses in our approach.

\section{Conclusions}
\label{sec:conclusions}
In this work, we revisit the calculation of medium-induced parton splitting and present a new analytic approach that allows for the first time to account for the various known limits. Our method, dubbed ``Improved Opacity expansion" resums multiple soft scatterings to all orders while treating single hard scattering as a perturbation.
To do so, following \cite{Mehtar-Tani:2019tvy}, we have suggested a shift of the expansion of the in-medium propagators around the so-called ``harmonic oscillator'' solution which takes into account diffusive momentum broadening. Perturbations around this solution correspond to hard, transverse kicks that reveal the quasi-particle structure of the underlying medium. In Ref.~\cite{Mehtar-Tani:2019tvy},  the radiative spectrum was calculated in the leading logarithmic approximation and therefore is applicable so long as $x_\perp \ll \mu^{-1}$, which translates into $\omega \gg \omega_\BH$. In the present work we go beyond by accounted for the full imaginary potential order by order. 
%

We have demonstrated the validity of the framework by computing the in-medium radiative spectrum. The NLO term, with a suitable choice of subtraction scale, allows to properly link the LPM regime, appropriate for dense media, to the regimes where single-scattering in the medium dominates, including the Bethe-Heitler regime at low frequencies, where formation time of the radiation probes the scale of the medium mean-free-path, and the GLV regime at high frequencies, where the formation time of the bremsstrahlung exceeds the length of the medium. This demonstrates that the approach, albeit formally suitable for large and dense media, where diffusive broadening dominate at small angles. 
The compactness of our main result and the good agreement with the results from an all-order resummation \cite{CaronHuot:2010bp}, which is implemented in the \texttt{MARTINI} event generator \cite{Schenke:2009gb,Park:2018acg}, makes it amenable for implementation in a fast Monte-Carlo event generator for jet quenching.

%



\acknowledgments
The authors thank Sangyong Jeon and Chanwook Park for providing numerical results for the code used in \cite{CaronHuot:2008ni}.
KT is supported by a Starting Grant from Trond Mohn Foundation (BFS2018REK01) and the University of Bergen. YMT is supported by the U.S. Department of Energy, Office of Science, Office of Nuclear Physics, under contract No. DE- SC0012704,
and in part by Laboratory Directed Research and Development (LDRD) funds from Brookhaven Science Associates.

\bibliographystyle{jhep}
\bibliography{bdmpsnlo}

\end{document}